\documentclass[notitlepage,nofootinbib,preprintnumbers,amssymb,superscriptaddress]{revtex4-1}
\usepackage{amsfonts,amssymb,amsmath,graphicx,color,bm}
\definecolor{ultramarine}{rgb}{0.07, 0.04, 0.56}
\definecolor{cadmiumgreen}{rgb}{0.0, 0.42, 0.24}
\definecolor{indigo(dye)}{rgb}{0.0, 0.25, 0.42}
\usepackage[linktocpage=true]{hyperref}
\usepackage[normalem]{ulem}
\hypersetup{
colorlinks=true,
citecolor=green,
linkcolor=cadmiumgreen,
urlcolor=indigo(dye),
}

\usepackage{autobreak}
\newcommand*{\D}{{\rm d}}

\def\[{\begin{equation}}
\def\]{\end{equation}}

\newcommand*{\cG}{{\mathcal{G}}}
\newcommand*{\cD}{{\mathcal{D}}}
\newcommand*{\cT}{{\mathcal{T}}}
\newcommand*{\Gt}{{\mathcal{G}_{\mathcal{T}}}}
\newcommand*{\Ft}{{\mathcal{F}_{\mathcal{T}}}}
\newcommand*{\tz}{\tilde{\zeta}}
\newcommand*{\dotp}{\dot{\pi}}
\newcommand*{\ddotp}{\ddot{\pi}}
\newcommand*{\be}{\begin{equation}}
\newcommand*{\ee}{\end{equation}}

\begin{document}

\title{Scalar-scalar-tensor interaction in DHOST theories}

\author{S. Mironov}
\email{sa.mironov\_1@physics.msu.ru}
\affiliation{Institute for Nuclear Research of the Russian Academy of Sciences,
60th October Anniversary Prospect, 7a, 117312 Moscow, Russia}
\affiliation{Institute for Theoretical and Mathematical Physics,
MSU, 119991 Moscow, Russia}
\affiliation{NRC, "Kurchatov Institute", 123182, Moscow, Russia}

\author{V. Volkova}
\email{volkova.viktoriya@physics.msu.ru}
\affiliation{Institute for Nuclear Research of the Russian Academy of Sciences,
60th October Anniversary Prospect, 7a, 117312 Moscow, Russia}
\affiliation{Department of Particle Physics and Cosmology, Physics Faculty, M.V. Lomonosov Moscow State University,
Vorobjevy Gory, 119991 Moscow, Russia}

\begin{abstract}
We derive the general cubic action for perturbations about a cosmological background in the quadratic subclass of degenerate higher-order scalar-tensor 
(DHOST) theories, focusing on scalar-scalar-tensor interactions. 
We adopt a fully covariant formulation and implement unitary gauge at the level of perturbations.
This mixed sector provides the key ingredient for estimating the decay rate of a gravitational wave into two scalar excitations in the quadratic DHOST models of dark energy in the late-time Universe.
\end{abstract}

\maketitle

\section{Introduction and summary}

The multi-messenger astronomy after the detection of the gravitational event GW170817~\cite{LIGOScientific:2017vwq} and its electromagnetic counterpart GRB170817A~\cite{Goldstein:2017mmi} has become a powerful probe for modifications to General Relativity (GR) in the late-time cosmology. 
In particular, the observational data has ruled out a significant number of modified gravity models with non-minimally coupled scalar fields, where
gravitational waves (GW) propagate through the cosmological medium at the speed $c_g$ that deviates from the speed of light $c$
by more than $10^{-15}$~\cite{LIGOScientific:2017zic}.
Such a stringent constraint on the GW propagation speed has considerably narrowed the class of (beyond) Horndeski theories~\cite{Horndeski:1974wa,Deffayet:2011gz,Gleyzes:2014dya,Gleyzes:2014qga} suitable for modelling dark energy (DE)~\cite{Sakstein:2017xjx,Ezquiaga:2017ekz,Creminelli:2017sry,Baker:2017hug,Amendola:2017orw,Langlois:2017dyl,Kase:2018aps}: 
{along with k-essence~\cite{Armendariz-Picon:2000ulo} and kinetic braiding~\cite{Deffayet:2010qz}
the exact luminality $c_{g} = c$ is ensured when the scalar field $\pi$
couples non-minimally to the Ricci scalar via $G_4(\pi)$ (see e.g.~\cite{Kase:2018aps} for a review). More general couplings typically lead to $c_{g} \neq c$, exceeding the observationally allowed deviation.
}

Recently there has been formulated an alternative viewpoint on constraining the scalar-tensor theories of modified gravity based on multi-messenger speed tests:
it was suggested in~\cite{KK1,Babichev:2024kfo,Mironov:2025dzz} to allow the scalar DE described by Horndeski theory to couple to the electromagnetic sector in a specific way so that $c_g/c = 1$. 
From a technical perspective, one of the possible ways to formulate the scalar-vector-tensor (SVT) theories of this kind 
is to make use of Kaluza-Klein compactification in a 5D Horndeski theory as it was shown in~\cite{KK1}. Indeed, in this case the respective couplings of the graviton and $U(1)$ gauge field to the DE scalar field automatically align, hence, ensuring the equality between the corresponding propagation speeds $c_g$ and $c$. The price to pay is that in this case the electromagnetic wave propagates at non-unit speed $c\neq 1$.
In result, it was shown that
the ratio $c_g/c = 1$ holds for a wider range of Horndeski subclasses than the multi-messenger speed tests allow in the case of $c=1$. 
What is more, this speeds' relation is valid for any cosmological background~\cite{KK1}, including the spherically symmetric one~\cite{Mironov:2024zzk}
\footnote{Let us note that imposing equality of the propagation speeds on an arbitrary background implies that this equality holds well beyond the redshift range $z$ directly probed by the data, see e.g.~\cite{Kase:2018aps}. This generalisation is reasonable, as it avoids fine-tuning and eliminates the need to model in detail the composition of the regions traversed by the signal.}. 
However, the analysis of~\cite{Babichev:2024kfo} showed that the scalar-photon couplings suggested 
in~\cite{KK1} can be undone by a disformal transformation of the metric so that in the resulting theory the electromagnetic field is minimally coupled to a disformal metric and, hence, viable
Horndeski theories are restricted back to the original subclass.
This subtle point with disformal equivalence was later taken into account in~\cite{KK_DHOST}, where a similar approach with compactification allowed to derive the scalar-photon couplings ensuring luminality of GW in  beyond Horndeski theories and in even more general class of degenerate higher-order scalar-tensor (DHOST) theories~\cite{Langlois:2015cwa,Crisostomi:2016czh,BenAchour:2016cay,BenAchour:2016fzp}. It appears that at least some the scalar-photon couplings are unremovable by disformal transformations of the metric, hence, constituting the "luminal" subclass of DHOST theories with $c_g/c = 1$.
{As a result, allowing for a non-minimal coupling to the electromagnetic sector has revived a considerable number of subclasses of Horndeski and DHOST theories as viable DE candidates, hence, providing more flexibility in confronting the growing body of observational data~\cite{DESI:2025zgx}.}

Another {type of constraints} for DE models based 
{on the scalar-tensor theories}
was discussed 
in~\cite{Creminelli:2018xsv,Creminelli:2019nok} and comes from the possible decay of the graviton into the DE {excitations}.
{The latter process becomes relevant due to 
the time-dependence of the background scalar field in scalar-tensor theories, which induces a preferred time-foliation of the FLRW metric. Hence,} Lorentz invariance is spontaneously broken and the propagation speed of the DE fluctuations differs from that of the massless graviton. In result, the decay of the graviton becomes kinematically allowed even for massive DE modes. 
It seems natural to assume that since the GW are observable on Earth, the possible decay rate should be negligible, which 
{allows to formulate }
an additional constraint on the {scalar-tensor theories in the late-time cosmology.} 
In~\cite{Creminelli:2018xsv} the decay rate for $g \to \pi\pi$ was derived within the Effective Field Theory (EFT) framework for those subclasses of Horndeski and beyond Horndeski theories where $c_g = c = 1$. The requirement for this decay to be suppressed provides a constraint for Lagrangian functions, which was taken into account in~\cite{KK_DHOST} when formulating the luminal beyond Horndeski subclass 
with $c_g/c = 1$ yet $c\neq 1$. 
So far the {suggested} luminal subclasses of DHOST theory in~\cite{KK_DHOST} comply with the limitations on the GW propagation speed only, while the 
{constraint} from non-decay of the GW is still missing for DHOST theories, at least according to our knowledge.

In this paper we aim to make a first step towards deriving such a constraint {for DHOST theories}
based on {the suppressed decay} of the GW.
Namely, the derivation of the decay rate for $g \to \pi\pi$ {requires 
computation of a mixed} scalar-scalar-tensor part of the cubic action for perturbations. In Sec.~\ref{sec:cubic_action} we present this mixed part of the cubic action for the most general quadratic DHOST theory in a fully covariant form. Where it is necessary, we specify to the most phenomenologically favoured DHOST Ia subclass, but only on the final stages of calculations. {For completeness, we also summarize the results} for the quadratic actions for both scalar and tensor perturbations in the quadratic DHOST theories in Sec.~\ref{sec:quadratic_action}. The main results of this paper are given {in eqs.~\eqref{eq:Lssh} and~\eqref{eq:Lssh_reduced}.}

\section{The theory and notations}

We focus on the subclass of DHOST theories, involving terms that are quadratic in $\nabla^2\pi$. 
Although these scalar-tensor theories involve higher derivative terms in the Lagrangian, by design they evade the Ostrogradsky ghost and describe $2+1$ dynamical degrees of freedom (DOFs) thanks to special relations between the Lagrangian functions usually referred to as degeneracy constraints, see e.g.~\cite{DHOSTReview} for a review.

The Lagrangian for the quadratic DHOST theory has the following general form 
(we adopt $(-,+,+,+)$ metric signature):
\begin{equation}
\begin{aligned}
\label{eq:lagrangian_DHOST}
\mathcal{L}_{DHOST_2} =  F(\pi,X) - K(\pi,X)\Box\pi + F_2(\pi,X) R + 
\sum_{i=1}^5 A_i(\pi,X) L_i \;, 
\qquad\qquad\qquad
\\
L_1 = \pi_{;\mu\nu} \pi^{;\mu\nu}, \quad
L_2= \left(\Box{\pi}\right)^2, \quad
L_3= \pi^{,\mu} \pi_{;\mu\nu} \pi^{,\nu} \Box{\pi}, \quad
L_4= \pi^{,\mu} \pi_{;\mu\nu} \pi^{;\nu\rho} \pi_{,\rho}, \quad
L_5= \left( \pi^{,\mu} \pi_{;\mu\nu} \pi^{,\nu} \right)^2,
\end{aligned}
\end{equation}
where $R$ is Ricci scalar, $\pi$ is a scalar field,
$X=-(g^{\mu\nu}\pi_{,\mu}\pi_{,\nu})/2$,
$\pi_{,\mu}=\partial_\mu\pi$,
$\pi_{;\mu\nu}=\nabla_\nu\nabla_\mu\pi$,
$\Box\pi = g^{\mu\nu}\nabla_\nu\nabla_\mu\pi$. 
Not all of the 8 scalar potentials $F$, $K$, $F_2$ and $A_i$ in~\eqref{eq:lagrangian_DHOST} are free, since, as we mentioned above, avoiding the Ostrogradsky ghost requires imposing certain relations on them. 
These degeneracy constraints subdivide the DHOST theories into subclasses, see e.g.~\cite{Crisostomi:2016czh,BenAchour:2016cay,BenAchour:2016fzp,DHOSTReview} 
for a complete classification. 
In what follows, up to some point we carry out calculations without imposing any relations on the scalar potentials, but at a certain step we focus on the so-called DHOST Ia subclass, which is considered to be the most phenomenologically sound. 
For DHOST Ia the potentials $A_2$, $A_4$ and $A_5$ 
are expressed through the independent
$F_2$, $A_1$ and $A_3$ as follows~\cite{DHOSTReview}
\footnote{Note the difference in notations: \cite{DHOSTReview} adopts $X=\pi_{,\mu} \pi^{,\mu}$ while in this paper  $X=-(\pi_{,\mu} \pi^{,\mu})/2$.}:
\begin{subequations}
  \label{eq:Ia_relations}
  \begin{align}
A_2 &= - A_1\,,\\
\label{a4_A}
A_4&= \frac{1}{8(F_2+ 2 X A_1)^2}\left[32 X A_1^3+4 (3F_2+16 XF_{2X})A_1^2
-4 X^2 F_2 A_3^2\qquad 
\right.
\cr
&\qquad 
\left.
- 8 ( 3XF_2 - 4 X^2 F_{2X}) A_3 A_1
+8 F_{2X}(3F_2+4XF_{2X})A_1
\right.
\cr
&\qquad 
\left.
+8F_2 (XF_{2X}-F_2)A_3+12F_2 F_{2X}^2\right],
\\
\label{a5_A}
A_5&=\frac{\left(F_{2X} + A_1 + X A_3\right)\left(A_1^2-3XA_1A_3+ F_{2X} A_1 - 2F_2 A_3\right)}{2(F_2 + 2XA_1)^2}\,,
  \end{align}
  \end{subequations}
  where $F_{2X} \equiv \partial F_2 / \partial X$, etc.
Other arbitrary potentials $F$ and $K$ belong to Horndeski subclass linear 
in $\nabla^2\pi$
and do not enter these relations. We will see shortly that the degeneracy relations~\eqref{eq:Ia_relations} indeed reduce the number of DOFs to only one scalar and two from tensor sector.

Let us recall that the quadratic DHOST theories~\eqref{eq:lagrangian_DHOST} involve quadratic Horndeski and beyond Horndeski theories as special cases.
To reproduce the results for the quadratic beyond Horndeski theories with the Lagrangian
\begin{eqnarray}
\label{eq:lagrangian_BH}
\mathcal{L}_{BH} &=& 
 F(\pi, X)-K(\pi, X)\Box\pi +G_4(\pi, X)R
\nonumber\\&&
+ G_{4X}\left[(\Box\pi)^2-(\nabla_\mu\nabla_\nu\pi)^2\right]
-F_4(\pi, X) 
\; \epsilon^{\mu\nu\rho}_{\sigma} \epsilon^{\mu'\nu'\rho'\sigma}
 \pi_{,\mu} \pi_{,\mu'} \pi_{;\nu\nu'} \pi_{;\rho\rho'},
\end{eqnarray}
one adopts the following choice for the potentials $A_i$ in~\eqref{eq:lagrangian_DHOST}
\footnote{Also note the difference in notations with~\cite{KK_DHOST}, where $X=\pi_{,\mu} \pi^{,\mu}$, $G_3$ corresponds to $(-K)$ and $F_4$ has the opposite sign as compared to this paper.}:
\begin{equation}
\label{eq:BH_limit}
F_2 = G_4, \quad A_2 = - A_1 = G_{4X} - 2 X F_4, \quad A_4 = - A_3 = 2 F_4, \quad A_5 = 0.
\end{equation}

\section{General form of the quadratic and cubic actions for cosmological perturbations}

In this section we present the general form of the quadratic and cubic actions for both scalar and tensor modes of perturbations over the cosmological background in the quadratic DHOST theories~\eqref{eq:lagrangian_DHOST}.

We consider a flat FLRW background with {a homogeneous scalar field $\pi = \pi(t)$} and employ the Arnowitt-Deser-Misner formalism for writing the metric as follows:
{ \be
\begin{aligned}
  \mathrm{d}s^2
= -N^2 \mathrm{d}t^2 +
\gamma_{ij}(\mathrm{d}x^i+ N^i \mathrm{d}t)(\mathrm{d}x^j
+ N^j \mathrm{d}t) \; ,
\qquad\qquad\qquad
\label{eq:ADM_metric}
\\
N= 1+\alpha, \qquad 
N_i = \partial_i\beta, \qquad
\gamma_{ij}= a^2(t) \; e^{2\zeta} \left(\delta_{ij} + h_{ij}^T +
\frac12 h_{ik}^T {h^{k\:T}_j}\right)\; ,
\end{aligned}
\ee
where 
$\alpha$, $\beta$ and $\zeta$ are lapse, shift and curvature perturbations, respectively, 
and $h_{ij}^T$ is traceless and
transverse tensor mode.} 
{We do not consider vector perturbations since in the scalar-tensor theories these modes are non-dynamical, see e.g.~\cite{KobaReview}.}
{We have completely fixed the gauge by removing the longitudinal scalar mode ($\partial_i\partial_j E = 0$) as well as the perturbations of the scalar field ($\delta \pi = 0$)}.
Since we are interested in both the quadratic and cubic actions
we consider both linear and quadratic perturbation terms in eq.~\eqref{eq:ADM_metric}, i.e. we use the following parametrization of the metric perturbations $\delta g_{\mu\nu}$:
\begin{equation}
\label{eq:metric_perturbation}
\begin{aligned}
	\delta g_{00} &= - (2 \alpha + \alpha^2) + {a^{-2}(t)} 
	\delta_{ij} \;\partial_i \beta \; \partial_j \beta, \\
	\delta g_{0i} &= \partial_i \beta, \\
	\delta g_{ij} &= a^2(t) \left[\delta_{ij} (2 \zeta + 2 \zeta^2) + h_{ij} (1 + 2 \zeta) + \frac12 h_{ik}\;h_{kj}\right].
\end{aligned}
\end{equation}
Note that we drop the terms that are cubic in perturbations, which is equivalent to making use of the background equations of motion (EOMs).

By expanding the Lagrangian~\eqref{eq:lagrangian_DHOST} up to the third order in metric perturbations $\delta g_{\mu\nu}$ we obtain the action for cosmological perturbations, which involves the quadratic and cubic contributions and reads schematically as follows:
\begin{equation}
\label{eq:quadratic_cubic}
\mathcal{S}^{(2+3)}=\int \D t\D^3x \left(
\mathcal{ L}_{hh}+\mathcal{ L}_{ss}+\mathcal{ L}_{hhh}
+\mathcal{ L}_{shh}+\mathcal{ L}_{ssh}+\mathcal{ L}_{sss}
\right).
\end{equation}
Although the quadratic parts $\mathcal{ L}_{hh}$ and $\mathcal{ L}_{ss}$
for DHOST theories were discussed in numerous existing works, see e.g.~\cite{Crisostomi:2018bsp} for the quadratic subclass,
we summarise the key results in the notations of this paper in Sec.~\ref{sec:quadratic_action}.

In this paper we focus on $\mathcal{ L}_{ssh}$ part which is the key for analysing the possible decay of the graviton into two scalars $\pi$.
The computation of $\mathcal{ L}_{ssh}$ in the quadratic DHOST theories~\eqref{eq:lagrangian_DHOST} within a covariant framework comprises the original result of this paper. As for the rest of the cubic contributions in~\eqref{eq:quadratic_cubic} we leave them for future studies.

\subsection{Quadratic action for the scalar and tensor modes}
\label{sec:quadratic_action}

The quadratic Lagrangian for the tensor perturbations $h_{ik}^T$ in the 
general quadratic DHOST theory~\eqref{eq:lagrangian_DHOST} reads:
\begin{equation}
\label{eq:Lhh}
\mathcal{ L}_{hh} = \frac{a^3}{8}\Bigg[{\mathcal{{G}_T}}\left(\dot{h}^T_{ik}\right)^2-\dfrac{\mathcal{F_T}}{a^2}\left(\partial_i h_{kl}^T\right)^2\Bigg],
\end{equation}
where
\begin{subequations}
\label{eq:tensor_coeff}
\begin{align}
\label{eq:GT_coeff_setup}
&\mathcal{G_T}=2 F_2 + 4 A_1 X,\\
&\mathcal{F_T}=2 F_2,
\end{align}
\end{subequations}
and in our notations $X = \dot{\pi}^2/2$, while the overdot indicates the derivative w.r.t. time $t$.
The GW propagation speed reads
\begin{equation}
\label{eq:GW_speed}
c_g^2 = \frac{\Ft}{\Gt}.
\end{equation} 

The quadratic Lagrangian for the scalar perturbations in the unitary gauge
($\delta \pi = 0$) reads:
\begin{eqnarray}
\label{eq:Lss}
\mathcal{ L}_{ss} &=& a^3\Bigg[
-3\left(\mathcal{{G}_T} + 3 \Lambda_2\right)\dot{\zeta}^2
+\mathcal{F_T}\dfrac{(\nabla\zeta)^2}{a^2}
-2(\mathcal{G_T}+\mathcal{D}) \;\alpha\dfrac{\nabla^2\zeta}{a^2}
+\Sigma\alpha^2
+6\Theta\alpha\dot{\zeta}
- 2\Theta\alpha\dfrac{\nabla^2\beta}{a^2}
\nonumber\\&&
+2\left(\mathcal{{G}_T} + 3 \Lambda_2\right)\dot{\zeta}\dfrac{\nabla^2\beta}{a^2} +
{\left(-\Lambda_3\dfrac{(\nabla\alpha)^2}{a^2} 
- \Lambda_1 \dot{\alpha} \dfrac{\nabla^2\beta}{a^2} 
+ 3\Lambda_1 \dot{\alpha}\dot{\zeta} 
+ \Lambda_4 \dot{\alpha}^2 
-\Lambda_2 \dfrac{(\nabla^2\beta)^2}{a^4}
\right)}
\Bigg],
\label{eq:quadr_action_DHOST}
\end{eqnarray}
where $(\nabla\zeta)^2 = \delta^{ij}\partial_i\zeta\partial_j\zeta$, $\nabla^2\beta = \delta^{ij}\partial_i\partial_j\beta$ and
\begin{subequations}
\label{eq:Scalar_coeff}
\begin{align}
\label{eq:D_coeff_setup}
&\mathcal{D}=-4 A_1 X - 4 F_{2X} X,\\
\label{eq:Theta_coeff_setup}
&\Theta=2 F_ 2 H + F_ {2\pi} \dot {\pi} - 6 A_ 1 H X - 12 A_ 2 H X - 
 2 F_ {2 X} H X + 2 F_ {2\pi X} X\dot {\pi} 
 \nonumber\\&
 - K_X X\dot {\pi} - 
 4 A_ {1 X} H X^2 - 12 A_ {2 X} H X^2 - 6 A_ 3 H X^2 + 
  \dot {\pi} \ddot {\pi} \left[A_ 1 - 2 A_ 2  \right.
 \nonumber\\&\left.
 + 3 F_ {2 X}  - 
 2 A_ {2 X} X
 + 
 3 A_ 3 X- 2 A_ 4 X + 
 2 F_ {2 XX} X
 + 
 2 A_ {3 X} X^2 + 
 4 A_ 5 X^2\right]
,\\
%
&\Lambda_1 = 4 X(-A_2 + F_{2X}  + A_3 X),\\
&\Lambda_2 = - 2 X (A_{1} + A_2),\\
&\Lambda_3 =  4 X (A_1 + F_{2X} -  A_4  X),\\
&\Lambda_4 = 2 X (A_ 1 + A_ 2 - 2 X (A_ 3 + A_ 4) + 4 A_ 5 X^2),
\end{align}
\end{subequations}
while the explicit form of $\Sigma$ is given in eq.~\eqref{eq:Sigma} in Appendix.
Let us note that $\mathcal{ L}_{ss}$ also involves $\alpha\zeta$ and $\zeta^2$ terms, which vanish upon using the background EOMs (see eqs.~\eqref{eq:EOM} in Appendix), hence, we omit them.
So far the results are given for the most general choice of the potentials $A_i$, i.e. without imposing the degeneracy conditions.
Note that the coefficients $\Lambda_i$ indentically vanish in (beyond) Horndeski theories, see the corresponding relations for the potentials~\eqref{eq:BH_limit}. 

At this point one has to make use of the degeneracy conditions to derive the quadratic Lagrangian for the sole scalar DOF.
Upon reducing the general case to the DHOST Ia subclass, i.e. imposing the degeneracy relations~\eqref{eq:Ia_relations}, we
immediately find that $\Lambda_2 = 0$, while $\Lambda_3$ and $\Lambda_4$ can be cast in the following form:
\begin{subequations}
\label{eq:Lambda_Ia}
\begin{align}
\Lambda_3 &= \frac{\Lambda_1 }{4 \Gt^2}\Big(\Lambda_1 \Ft + 4 \Gt (\Gt+\cD)\Big), \\
\Lambda_4 &= -\frac{3 \Lambda_1^2}{4 \cG_\cT}. 
\end{align}
\end{subequations}
Then, as expected, 
the terms $\dot{\zeta}^2$, $\dot{\alpha}^2$ and $\dot\zeta \dot\alpha$ in~\eqref{eq:Lss} get combined into a perfect square, i.e. the corresponding kinetic matrix is indeed degenerate. 
The latter allows one to make a field redefinition
\[
\label{eq:field_redef}
	\tz = \zeta - \Delta_1\alpha, \qquad\mbox{with} \qquad \Delta_1 = \frac{\Lambda_1}{2\Gt},
\]
so that the quadratic Lagrangian~\eqref{eq:quadr_action_DHOST} features only $\alpha$, $\beta$ and $\tz$:
\[
\begin{aligned}
\label{eq:Lss_newzeta}
\mathcal{L}_{ss} &= 
a^3\Bigg[
-3\mathcal{{G}_T}\dot{\tz}^2 +\mathcal{F_T}\dfrac{(\nabla\tz)^2}{a^2}
-2(\mathcal{G_T}+\mathcal{D}+
\mathcal{F_T}\Delta_1)\alpha\dfrac{\nabla^2\tz}{a^2} \\ 
&+ \tilde{\Sigma}\alpha^2
+ 6 \tilde{\Theta} \alpha\dot{\tz}
-2 \tilde{\Theta} \alpha\dfrac{\nabla^2\beta}{a^2}
+2\mathcal{{G}_T}\dot{\tz}\dfrac{\nabla^2\beta}{a^2}
\Bigg],
\end{aligned}
\]
{where  
\begin{subequations}
\label{eq:App_theta_sigma}
\begin{align}
&\tilde{\Theta} = \Theta - \mathcal{G_T}\dot{\Delta}_1, \\
&\tilde{\Sigma} = \Sigma + 3\mathcal{G_T} \dot{\Delta}_1^2+ 6\tilde{\Theta} \dot{\Delta}_1 
-\dfrac{3}{a^3}\dfrac{\mathrm{d}}{\mathrm{d}t}\Big[a^3\left(\tilde{\Theta} +\mathcal{G_T} \dot{\Delta}_1\right)\Delta_1\Big].
\end{align}
\end{subequations}
Note that the coefficients $\tilde{\Theta}$ and $\tilde{\Sigma}$ play the similar role to $\Theta$ and $\Sigma$ in (beyond) Horndeski theory since $\Delta_1$ vanishes in this limit, see e.g.~\cite{KobaReview}. 

Varying eq.~\eqref{eq:Lss_newzeta} w.r.t. $\alpha$ and $\beta$ gives two constraint equations:
\begin{equation}
\label{eq:quadr_contraints}
\begin{aligned}
\alpha &= \frac{\mathcal{G_T}}{\tilde\Theta} \dot{\tz},
\\
\dfrac{\nabla^2\beta}{a^2} &= 
\frac{1}{\tilde\Theta} \left( \left[\frac{\tilde{\Sigma}\Gt}{\tilde\Theta} + 3 \tilde{\Theta}\right] \dot{\tz}
 - \left(\mathcal{G_T} + \mathcal{D} + \mathcal{F_T}\Delta_1 \right) 
\dfrac{\nabla^2\tz}{a^2} \right),
\end{aligned}
\end{equation}
Then upon substituting these constraints to~\eqref{eq:Lss_newzeta} we arrive to the final form of the quadratic Lagrangian in DHOST Ia theory which features the only dynamical scalar DOF $\tz$:
\begin{equation}
\label{eq:quadr_action_DHOSTIa-scalar-fin}
\begin{aligned}
  \mathcal{L}_{ss}=
  a^3 \Bigg[ \mathcal{{G}_S} \dot{\tilde{\zeta}}^2
- \dfrac{1}{a^2}  \mathcal{{F}_S} (\partial_i \tilde{\zeta})^2
\Bigg] \; ,
\end{aligned}
\end{equation}
where
\begin{subequations}
\label{eq:GsFs}
\begin{align}
  \mathcal{G_S} & = \dfrac{\tilde{\Sigma}\mathcal{{G}_T}^2}{\tilde{\Theta}^2}
    +3\mathcal{{G}_T}, \label{eq:Gs}
  \\
  \mathcal{F_S} &= \dfrac{1}{a}\dfrac{\mathrm{d}}{\mathrm{d}t}
  \left[ \dfrac{a \;\mathcal{{G}_T}\left(\mathcal{G_T} + \mathcal{D}  + \mathcal{F_T}\Delta_1\right)}{\tilde{\Theta}}\right]
  -\mathcal{F_T}.
   \label{eq:Fs}
\end{align}
\end{subequations}
The propagation speed of the scalar mode $\tz$ reads:
\begin{equation}
\label{eq:scalar_speed}
c_{\mathcal{S}}^2 = \frac{\mathcal{F_S}}{\mathcal{G_S}}.
\end{equation}
As before one can immediately restore the existing results in both Horndeski and beyond Horndeski limits by taking $\Delta_1 \to 0$, see~\cite{KobaReview}.
We note in passing that the form of the propagations speeds in eqs.~\eqref{eq:GW_speed} and~\eqref{eq:scalar_speed} justify the possibility of the graviton decay in theories like~\eqref{eq:lagrangian_DHOST}, since we 
consider $c_g^2 \neq 1$ and generally $c_{\mathcal{S}}^2 \neq c_g^2$, i.e. the decay is kinematically allowed {even for massive scalar modes.}

\subsection{Scalar-scalar-tensor interactions in the cubic action}
\label{sec:cubic_action}

In this section we present the cubic Lagrangian for the most general quadratic DHOST theory~\eqref{eq:lagrangian_DHOST} featuring the interactions of two scalars and one tensor mode: 
\begin{eqnarray}
\label{eq:Lssh}
\mathcal{ L}_{ssh}&=&a\left[
2\Theta \alpha\beta_{,ij}h_{ij} +\frac{\Gamma }{2}
\alpha\beta_{,ij}\dot h_{ij}
+ \Lambda_1 \dot{\alpha}\beta_{,ij} h_{ij}
-\frac{3\mathcal{ G}_T}{2}\zeta\beta_{,ij}\dot h_{ij}
-2(\mathcal{ G}_T + 3 \Lambda_2)\dot\zeta\beta_{,ij}h_{ij}
\right.
\nonumber\\&&
\left.
-\mathcal{ F}_T\zeta_{,i}\zeta_{,j}h_{ij}+2(\mathcal{ G}_T+\mathcal{D})\alpha\zeta_{,ij}h_{ij}
+\frac{\mathcal{ G}_T}{2a^2}\beta_{,ij}\beta_{,k}h_{ij,k}
+ \frac{2\Lambda_2}{a^2} \beta_{,ij}\beta_{,kk} h_{ij}
+ \Lambda_3 \alpha_{,i}\alpha_{,j} {h}_{ij}
\right],
\end{eqnarray}
where we dropped the superscript in $h_{ij}^T$ and adopted shorter notations
$\beta_{,ij} \equiv \partial_i\partial_j \beta$ for brevity, 
and introduced a new variable
\begin{equation}
\Gamma= 2 F_2 + 12 A_1 X + 4 F_{2X} X + 8 A_{1X} X^2.
\end{equation}
Let us recall that $\Lambda_1$, $\Lambda_2$  and $\Lambda_3$ are characteristic coefficients for the DHOST theory and identically vanish upon taking the beyond Horndeski limit~\eqref{eq:BH_limit}, see eqs.~\eqref{eq:Scalar_coeff}, hence, restoring 
the existing results for $\mathcal{L}_{ssh}$ in Horndeski 
theory~\cite{Gao:2012ib} and beyond Horndeski theory~\cite{Creminelli:2018xsv}.
Let us note that the formerly valid relation $\Gamma = \partial\Theta/\partial H$ in (beyond) Horndeski case no more holds in DHOST theories. 
Finally, we note for completeness
that the action~\eqref{eq:Lssh} also features structures like $\zeta \beta_{,ij} h_{ij}$,
$\beta_{,i}\beta_{,j}\dot{h}_{ij}$, which vanish automatically, and
$\beta_{,i}\beta_{,j}{h}_{ij}$, which is vanishing upon making use of the background EOMs.

The further step is to impose DHOST Ia degeneracy constraints~\eqref{eq:Ia_relations} in the cubic Lagrangian $\mathcal{L}_{ssh}$~\eqref{eq:Lssh}, which, in particular, removes $\Lambda_2$ term and relates $\Lambda_3$ to other variables in~\eqref{eq:tensor_coeff} and~\eqref{eq:Scalar_coeff}, see eq.~\eqref{eq:Lambda_Ia}. 
{Then in order to make use of the constraints~\eqref{eq:quadr_contraints} 
for integrating out non-dynamical $\alpha$ and $\beta$ in the Lagrangian~\eqref{eq:Lssh}, we have to first adopt the field 
redefinition~\eqref{eq:field_redef}. The result reads: 
\begin{eqnarray}
\label{eq:Lssh_DHOSTIa}
\mathcal{ L}_{ssh}&=&a\left[
2\tilde{\Theta} \alpha\beta_{,ij}h_{ij} 
+\frac12 \left({\Gamma }-3 \mathcal{G}_T \Delta_1\right)
\alpha\beta_{,ij}\dot h_{ij}
-\frac{3\mathcal{ G}_T}{2}\tz\beta_{,ij}\dot h_{ij}
-2\mathcal{ G}_T \dot\tz\beta_{,ij}h_{ij}
-\mathcal{ F}_T\tz_{,i}\tz_{,j}h_{ij}
\right.
\nonumber\\&&
\left.
+2(\mathcal{ G}_T+\mathcal{D} + \mathcal{F}_T \Delta_1)\alpha\tz_{,ij}h_{ij}
+\frac{\mathcal{ G}_T}{2a^2}\beta_{,ij}\beta_{,k}h_{ij,k}
\right],
\end{eqnarray}
where we use the notations for coefficients $\tilde{\Theta}$ and $\Delta_1$ introduced in the quadratic Lagrangian, see eqs.~\eqref{eq:field_redef} and~\eqref{eq:App_theta_sigma}.
Substituting $\alpha$ and $\beta$ in terms of $\dot\tz$, $\tz$ and $\psi \equiv \nabla^{-2} \dot\tz$ from
the constraint equations~\eqref{eq:quadr_contraints}, we obtain the reduced form of the cubic Lagrangian:
\begin{eqnarray}
\label{eq:Lssh_reduced}
\mathcal{ L}_{ssh}&=&a^3\left[\frac{c_1}{a^2}
h_{ij}\tz_{,i}\tz_{,j}
+\frac{c_2}{a^2}\dot h_{ij}\tz_{,i}\tz_{,j}
+c_3\dot h_{ij}\tz_{,i}\psi_{,j}
+\frac{c_4}{a^2}\partial^2h_{ij}\tz_{,i}\psi_{,j}
\right.
\nonumber\\&&
\left.
+\frac{c_5}{a^4}\partial^2 h_{ij}\tz_{,i}\tz_{,j}
+c_6\partial^2 h_{ij}\psi_{,i}\psi_{,j}
+ \frac{c_7}{a^2} \dot h_{ij}\dot\tz_{,i}\tz_{,j} 
+c_8 \dot{h}_{ij} \dot\tz_{,i} \psi_{,j} \right],
\end{eqnarray}
where $\partial^2 h_{ij} \equiv h_{ij,kk}$ and the coefficients read
\begin{align}
c_1 &= \mathcal{F}_S, & \qquad 
c_5 &= \frac{\mathcal{G}_T}{4 \tilde{\Theta}^2} (\mathcal{G}_T +\mathcal{D} +\mathcal{F}_T \Delta_1)^2, \\
c_2 &= -\frac{\mathcal{G}_T}{2 \tilde{\Theta}} (\mathcal{G}_T +\mathcal{D} +\mathcal{F}_T \Delta_1), & \qquad 
c_6 &= \frac{\mathcal{G}_S^2}{4\mathcal{G}_T}, \\
c_3 &= \frac32 \mathcal{G}_S, & \qquad 
c_7 &= \frac{\mathcal{G}_T(\Gamma - 3 \mathcal{G}_T \Delta_1)}{2\tilde{\Theta}^2} (\mathcal{G}_T +\mathcal{D} +\mathcal{F}_T \Delta_1),\\
c_4 &= -\frac{\mathcal{G}_S}{2 \tilde{\Theta}} (\mathcal{G}_T +\mathcal{D} +\mathcal{F}_T \Delta_1), & \qquad 
c_8 &= -\frac{\mathcal{G}_S(\Gamma - 3 \mathcal{G}_T \Delta_1)}{2\tilde{\Theta}}.
\end{align}
Note that the reduced form of $L_{ssh}$~\eqref{eq:Lssh_reduced} involves solely the tensor mode $h_{ij}$ and the only scalar DOF $\tz$ and holds for any DHOST Ia subclass. Upon taking the Horndeski limit (i.e. $\Delta_1 \to 0$, $\mathcal{D} \to 0$) in the Lagrangian~\eqref{eq:Lssh_reduced} one reproduces the existing result given in~\cite{Gao:2012ib}
\footnote{It appears that there is typo in~\cite{Gao:2012ib}: the first term in eq.(61) is likely to feature $\zeta_{,j} \dot{h}_{ij}$ 
{instead of the original $\zeta_{,j} {h}_{ij}$, so the expression should read 
$\bar f_i:=\frac{\Gamma}{2\Theta}\zeta_{,j}\dot{h}_{ij}
+\frac{\mu}{\mathcal{ G}_T}\zeta_{,j}\dot h_{ij}
+\frac{\mu}{a^2\Theta}\zeta_{,j}\partial^2 h_{ij}
-\frac{\mu\mathcal{ G}_S}{\mathcal{ G}_T^2}\psi_{,j}\partial^2 h_{ij}$. }}. 

The cubic Lagrangian~\eqref{eq:Lssh_reduced}  is the key part for 
computation of the decay rate for $g \to \pi\pi$ and an associated constraint for the potentials $F_2$ and $A_i$. We leave the corresponding derivations for the near future.
}
\section*{Acknowledgements}
The work 
of S.M. and V.V. has been supported by Theoretical Physics and Mathematics Advancement Foundation “BASIS”.


\section*{Appendix}
\label{sec:app}

In this Appendix we provide the explicit form of the coefficient $\Sigma$ 
appearing in the quadratic Lagrangian~\eqref{eq:Lss} for scalar modes in the quadratic DHOST theory~\eqref{eq:lagrangian_DHOST}:
\begin{equation}
\label{eq:Sigma}
\begin{aligned}
&\Sigma = F_X X + 2 F_{XX} X^2 - 2 K_{\pi} X - 2 K_{\pi X} X^2 + 6 H X\dotp (2K_X +K_{XX}X)
\\
&-6 H^2 F_2 + 6 H^2 X (7 F_{2X} + 4 F_{2XX} X) - 6 H \dotp (F_{2\pi} + F_{2\pi X} X + (2 F_{2X} + F_{2XX} X)\ddotp)
\\
 &+ 12 \dot{H} X (2F_{2X} + F_{2XX} X) 
+ 6 H^2 X (6 A_1 + X(9 A_{1X} + 2 A_{1XX} X)) - 6 H(2 A_1 + A_{1X} X)\dotp\ddotp
\\
&-4 X (2 A_{1\pi} + A_{1\pi X} X)\ddotp + (2A_1 - X(5 A_{1X} + 2 A_{1XX} X))\ddotp^2 - 2 (2 A_1 + A_{1X} X) \dotp \dddot{\pi}
\\
&+18 H^2 X (2 A_2 +X (7 A_{2X} + 2 A_{2XX} X)) - 12 H X \dotp(2 A_{2\pi} +A_{2\pi X}X)
\\
&- 12 \dot{H} X (2 A_2 +A_{2X} X) - 4 X (2 A_{2\pi} + A_{2\pi X} X) \ddotp 
+ (2 A_2 - 5 A_{2X}X - 2 A_{2XX}X^2)\ddotp^2 
\\
&- 2 (2 A_2 + A_{2X}X)\dotp\dddot{\pi}+ 36 H^2 X^2 (3 A_3 + A_{3X} X) + 6 H X \dotp (2X(3 A_{3\pi} + A_{3\pi X}X) + (3 A_3 + A_{3X} X)\ddotp)
\\
&+ 12 \dot{H} X^2 (3 A_3 + A_{3X} X) +8 X^2 (3 A_{3\pi} + A_{3\pi X}X)\ddotp
+ 2 X (3 A_3 + 9 A_{3X} X + 2 A_{3XX} X^2) \ddotp^2 
\\
&+ 4 X (3 A_3 + A_{3X} X) \dotp \dddot{\pi}
+ 12 H X \dotp \ddotp (3 A_4 + A_{4X}X) 
+ 8 X^2 (3 A_{4\pi} + A_{4\pi X} X)\ddotp 
\\
&+ 2 X (3 A_4 +X(9 A_{4X} + 2 A_{4XX}X))\ddotp^2 + 4 X (3 A_4 +A_{4X}X)\dotp \dddot{\pi}- 24 H X^2 (4 A_5 +A_{5X} X)\dotp\ddotp 
\\
&-16 X^3 (4 A_{5\pi} + A_{5\pi X}X)\ddotp -4 X^2 (12 A_5 +X(13 A_{5X} + 2 A_{5XX} X))\ddotp^2 
- 8 X^2 (4 A_5 +A_{5X}X)\dotp \dddot{\pi}.
\end{aligned}
\end{equation}
We also provide the background EOMs for the quadratic DHOST theory~\eqref{eq:lagrangian_DHOST} for completeness:
\begin{subequations}
\label{eq:EOM}
\begin{align}
\delta g^{00}: \quad  
&F - 2 F_X X - 6 H K_X X \dotp + 2 K_{\pi} X
+ 6 H^2 (F_2 - 4 F_{2X} X) + 6 H \dotp (F_{2\pi}  + F_{2X}\ddotp) 
\\\nonumber
&
-12 \dot{H} F_{2X} X - 6 H^2 X (3 A_1 +2 A_{1X} X) + 6 H A_1 \dotp\ddotp
+4 A_{1\pi} X \ddotp - (A_1-2A_{1X} X)\ddotp^2 
\\\nonumber
& + 2A_1 \dotp \dddot{\pi} -18 H^2 X (A_2 + 2 A_{2X} X) + 12 H A_{2\pi} X \dotp + 12 \dot{H} A_2 X
+4 A_{2\pi} X \ddotp 
\\\nonumber
&- (A_2 - 2 A_{2X} X) \ddotp^2 +2 A_2 \dotp \dddot{\pi}
-36 H^2 A_3 X^2 - 6 H X \dotp (2 A_{3\pi} X + A_3 \ddotp) 
\\\nonumber
&- 2 X (6 \dot{H} A_3 X + 4 A_{3\pi} X \ddotp + (A_3+2A_{3X} X)\ddotp^2 +2 A_3 \dotp \dddot{\pi})
\\\nonumber
&- 12 H A_4 X \dotp\ddotp - 2 X (4 A_{4\pi} X \ddotp + (A_4+2 A_{4X} X) \ddotp^2 +2 A_4 \dotp \dddot{\pi})
\\\nonumber
&+ 24 H A_5 X^2 \dotp \ddotp + 4 X^2 (4 A_{5\pi} X \ddotp + (3 A_5 +2 A_{5X} X)\ddotp^2 + 2 A_5 \dotp \dddot{\pi}) = 0,
\\
\delta g^{ij}: \quad 
&F - 2 X ( K_{\pi} + K_X \ddotp) + 6 H^2 F_2 + 4 H \dotp (F_{2\pi} +F_{2X} \ddotp)  
\\\nonumber
&+ 2 (2 \dot{H} F_2 + 2 F_{2\pi\pi} X + (F_{2\pi} + 4 F_{2\pi X} X) \ddotp + (F_{2X} + 2 F_{2XX} X)\ddotp^2 + F_{2X} \dotp\dddot{\pi}) 
\\\nonumber
&- 6 H^2 A_1 X - 4 H \dotp (A_{1\pi} X + (A_1+A_{1X} X)\ddotp) - 4\dot{H}A_1 X + A_1 \ddotp^2
\\\nonumber
&- 18 H^2 A_2 X - 12 H \dotp (A_{2\pi} X + (A_2 + A_{2X}X)\ddotp)
- 12 \dot{H} A_2 X - 4A_{2\pi} X \ddotp 
\\\nonumber
&- (A_2+4A_{2X} X)\ddotp^2 - 2A_2 \dotp\dddot{\pi}
+ 2 X (2 A_{3\pi} X \ddotp + 2 (A_3+A_{3X}X)\ddotp^2 +A_3 \dotp\dddot{\pi})
\\\nonumber
&- 2 X \ddotp^2 (A_4 - 2 A_5 X) = 0
.
\end{align}
\end{subequations}



\begin{thebibliography}{99}
{\small


\bibitem{LIGOScientific:2017vwq}
B.~P.~Abbott \textit{et al.} [LIGO Scientific and Virgo],
``GW170817: Observation of Gravitational Waves from a Binary Neutron Star Inspiral,''
Phys. Rev. Lett. \textbf{119} (2017) no.16, 161101
[arXiv:1710.05832 [gr-qc]].

\bibitem{Goldstein:2017mmi}
A.~Goldstein, P.~Veres, E.~Burns, M.~S.~Briggs, R.~Hamburg, D.~Kocevski, C.~A.~Wilson-Hodge, R.~D.~Preece, S.~Poolakkil and O.~J.~Roberts, \textit{et al.}
``An Ordinary Short Gamma-Ray Burst with Extraordinary Implications: Fermi-GBM Detection of GRB 170817A,''
Astrophys. J. Lett. \textbf{848} (2017) no.2, L14
[arXiv:1710.05446 [astro-ph.HE]].

\bibitem{LIGOScientific:2017zic}
B.~P.~Abbott \textit{et al.} [LIGO Scientific, Virgo, Fermi-GBM and INTEGRAL],
``Gravitational Waves and Gamma-rays from a Binary Neutron Star Merger: GW170817 and GRB 170817A,''
Astrophys. J. Lett. \textbf{848} (2017) no.2, L13
[arXiv:1710.05834 [astro-ph.HE]].


\bibitem{Horndeski:1974wa}
G.~W.~Horndeski,
``Second-order scalar-tensor field equations in a four-dimensional space,''
Int. J. Theor. Phys. \textbf{10} (1974), 363-384

\bibitem{Deffayet:2011gz}
C.~Deffayet, X.~Gao, D.~A.~Steer and G.~Zahariade,
``From k-essence to generalised Galileons,''
Phys. Rev. D \textbf{84} (2011), 064039
[arXiv:1103.3260 [hep-th]].


\bibitem{Gleyzes:2014dya}
J.~Gleyzes, D.~Langlois, F.~Piazza and F.~Vernizzi,
``Healthy theories beyond Horndeski,''
Phys. Rev. Lett. \textbf{114} (2015) no.21, 211101
[arXiv:1404.6495 [hep-th]].

\bibitem{Gleyzes:2014qga}
J.~Gleyzes, D.~Langlois, F.~Piazza and F.~Vernizzi,
``Exploring gravitational theories beyond Horndeski,''
JCAP \textbf{02} (2015), 018
[arXiv:1408.1952 [astro-ph.CO]].


\bibitem{Sakstein:2017xjx}
J.~Sakstein and B.~Jain,
``Implications of the Neutron Star Merger GW170817 for Cosmological Scalar-Tensor Theories,''
Phys. Rev. Lett. \textbf{119} (2017) no.25, 251303
[arXiv:1710.05893 [astro-ph.CO]].

\bibitem{Ezquiaga:2017ekz}
J.~M.~Ezquiaga and M.~Zumalac{\'a}rregui,
``Dark Energy After GW170817: Dead Ends and the Road Ahead,''
Phys. Rev. Lett. \textbf{119} (2017) no.25, 251304
[arXiv:1710.05901 [astro-ph.CO]].


\bibitem{Creminelli:2017sry}
P.~Creminelli and F.~Vernizzi,
``Dark Energy after GW170817 and GRB170817A,''
Phys. Rev. Lett. \textbf{119} (2017) no.25, 251302
[arXiv:1710.05877 [astro-ph.CO]].

\bibitem{Baker:2017hug}
T.~Baker, E.~Bellini, P.~G.~Ferreira, M.~Lagos, J.~Noller and I.~Sawicki,
``Strong constraints on cosmological gravity from GW170817 and GRB 170817A,''
Phys. Rev. Lett. \textbf{119} (2017) no.25, 251301
[arXiv:1710.06394 [astro-ph.CO]].

\bibitem{Amendola:2017orw}
L.~Amendola, M.~Kunz, I.~D.~Saltas and I.~Sawicki,
``Fate of Large-Scale Structure in Modified Gravity After GW170817 and GRB170817A,''
Phys. Rev. Lett. \textbf{120} (2018) no.13, 131101
[arXiv:1711.04825 [astro-ph.CO]].

\bibitem{Langlois:2017dyl}
D.~Langlois, R.~Saito, D.~Yamauchi and K.~Noui,
``Scalar-tensor theories and modified gravity in the wake of GW170817,''
Phys. Rev. D \textbf{97} (2018) no.6, 061501
[arXiv:1711.07403 [gr-qc]].



\bibitem{Kase:2018aps}
R.~Kase and S.~Tsujikawa,
``Dark energy in Horndeski theories after GW170817: A review,''
Int. J. Mod. Phys. D \textbf{28} (2019) no.05, 1942005
[arXiv:1809.08735 [gr-qc]].

\bibitem{Armendariz-Picon:2000ulo}
C.~Armendariz-Picon, V.~F.~Mukhanov and P.~J.~Steinhardt,
``Essentials of k essence,''
Phys. Rev. D \textbf{63} (2001), 103510
[arXiv:astro-ph/0006373 [astro-ph]].

\bibitem{Deffayet:2010qz}
C.~Deffayet, O.~Pujolas, I.~Sawicki and A.~Vikman,
``Imperfect Dark Energy from Kinetic Gravity Braiding,''
JCAP \textbf{10} (2010), 026
[arXiv:1008.0048 [hep-th]].



\bibitem{KK1}
S.~Mironov, A.~Shtennikova and M.~Valencia-Villegas,
``Reviving Horndeski after GW170817 by Kaluza-Klein compactifications,''
Phys. Lett. B \textbf{858} (2024), 139058
[arXiv:2405.02281 [hep-th]].

\bibitem{Babichev:2024kfo}
E.~Babichev, C.~Charmousis, B.~Muntz, A.~Padilla and I.~D.~Saltas,
``Horndeski speed tests with scalar-photon couplings,''
JCAP \textbf{01} (2025), 041
[arXiv:2407.20339 [gr-qc]].

\bibitem{Mironov:2025dzz}
S.~Mironov, A.~Shtennikova and M.~Valencia-Villegas,
``Ghost-free, gauge invariant SVT generalizations of Horndeski theory,''
Eur. Phys. J. C \textbf{85} (2025) no.12, 1378
[arXiv:2509.16850 [hep-th]].

\bibitem{Mironov:2024zzk}
S.~Mironov, M.~Sharov and V.~Volkova,
``Time-dependent, spherically symmetric background in Kaluza-Klein compactified Horndeski theory and the speed of gravity waves,''
JCAP \textbf{09} (2025), 047
[arXiv:2408.06329 [gr-qc]].


\bibitem{KK_DHOST}
S.~Mironov, A.~Shtennikova and M.~Valencia-Villegas,
``Luminal scalar-tensor theories for a not so dark dark energy,''
Phys. Rev. D \textbf{111} (2025) no.10, 10
[arXiv:2412.13460 [hep-th]].


\bibitem{Langlois:2015cwa}
D.~Langlois and K.~Noui,
``Degenerate higher derivative theories beyond Horndeski: evading the Ostrogradski instability,''
JCAP \textbf{02} (2016), 034
[arXiv:1510.06930 [gr-qc]].

\bibitem{Crisostomi:2016czh}
M.~Crisostomi, K.~Koyama and G.~Tasinato,
``Extended Scalar-Tensor Theories of Gravity,''
JCAP \textbf{04} (2016), 044
[arXiv:1602.03119 [hep-th]].

\bibitem{BenAchour:2016cay}
J.~Ben Achour, D.~Langlois and K.~Noui,
``Degenerate higher order scalar-tensor theories beyond Horndeski and disformal transformations,''
Phys. Rev. D \textbf{93} (2016) no.12, 124005
[arXiv:1602.08398 [gr-qc]].

\bibitem{BenAchour:2016fzp}
J.~Ben Achour, M.~Crisostomi, K.~Koyama, D.~Langlois, K.~Noui and G.~Tasinato,
``Degenerate higher order scalar-tensor theories beyond Horndeski up to cubic order,''
JHEP \textbf{12} (2016), 100
[arXiv:1608.08135 [hep-th]].

\bibitem{DESI:2025zgx}
M.~Abdul Karim \textit{et al.} [DESI],
``DESI DR2 results. II. Measurements of baryon acoustic oscillations and cosmological constraints,''
Phys. Rev. D \textbf{112} (2025) no.8, 083515
[arXiv:2503.14738 [astro-ph.CO]].

\bibitem{Creminelli:2018xsv}
P.~Creminelli, M.~Lewandowski, G.~Tambalo and F.~Vernizzi,
``Gravitational Wave Decay into Dark Energy,''
JCAP \textbf{12} (2018), 025
[arXiv:1809.03484 [astro-ph.CO]].

\bibitem{Creminelli:2019nok}
P.~Creminelli, G.~Tambalo, F.~Vernizzi and V.~Yingcharoenrat,
``Resonant Decay of Gravitational Waves into Dark Energy,''
JCAP \textbf{10} (2019), 072
[arXiv:1906.07015 [gr-qc]].



\bibitem{DHOSTReview}
D.~Langlois,
 ``Dark energy and modified gravity in degenerate higher-order scalar\textendash{}tensor (DHOST) theories: A review,''
Int. J. Mod. Phys. D \textbf{28} (2019), 1942006
[arXiv:1811.06271 [gr-qc]].

\bibitem{Crisostomi:2018bsp}
M.~Crisostomi, K.~Koyama, D.~Langlois, K.~Noui and D.~A.~Steer,
``Cosmological evolution in DHOST theories,''
JCAP \textbf{01} (2019), 030
[arXiv:1810.12070 [hep-th]].


\bibitem{KobaReview}
  T.~Kobayashi,
  ``Horndeski theory and beyond: a review,''
  Rept.\ Prog.\ Phys.\  {\bf 82} (2019),  086901
[arXiv:1901.07183 [gr-qc]].

\bibitem{Gao:2012ib}
X.~Gao, T.~Kobayashi, M.~Shiraishi, M.~Yamaguchi, J.~Yokoyama and S.~Yokoyama,
``Full bispectra from primordial scalar and tensor perturbations in the most general single-field inflation model,''
PTEP \textbf{2013} (2013), 053E03
[arXiv:1207.0588 [astro-ph.CO]].

}
\end{thebibliography}
\end{document}